*Article*

# Optimal Demand Response-Based Energy Management for Campus Buildings with Battery Energy Storage Systems: A Case Study of the ELB Building

Ali Beidoun[1], Quang Manh Hoang[1], Guilherme Vieira Hollweg[1], and Van-Hai Bui[1,*]

1 Department of Electrical and Computer Engineering, University of Michigan-Dearborn, USA; alibeido@umich.edu (A.B.), manhhq@umich.edu (Q.M. H.), hollweg@umich.edu (G.V. H.), vhbui@umich.edu (V.-H. B)
* Correspondence: vhbui@umich.edu (V.-H. B); Tel.: +1 734 686 1940.

**Abstract**

Buildings account for a significant portion of global electricity consumption, making demand response (DR) an effective strategy for reducing energy costs and improving grid flexibility. This paper presents an optimal energy management framework for a campus building that integrates demand response and battery energy storage systems (BESS) to minimize electricity costs while satisfying operational constraints and occupant comfort requirements. The proposed optimization model schedules building loads and battery charging/discharging based on time-varying electricity prices and building demand profiles. The framework is validated using the Engineering Laboratory Building (ELB) at the University of Michigan-Dearborn as a real-world case study. Four operating scenarios are investigated: (i) baseline operation, (ii) demand response without battery energy storage, (iii) battery energy storage, and (iv) demand response with battery energy storage. Simulation results demonstrate that demand alone reduces the building's electricity cost by approximately 12% compared with conventional operation. When battery energy storage is incorporated, the total electricity cost reduction increases to approximately 26%, highlighting the significant economic benefits of coordinated demand response and energy storage scheduling. The results demonstrate that integrating battery energy storage with demand response substantially enhances the operational flexibility of campus buildings and provides an effective solution for cost-effective and sustainable building energy management.

## 1. Introduction

Buildings constitute a significant portion of global energy consumption, making improvements in building energy efficiency, operational control, and demand-side flexibility increasingly important for reducing operating costs and supporting more sustainable power systems [1–5]. Large commercial and institutional buildings are particularly relevant because their electricity demand varies with HVAC operation, lighting, occupancy, laboratory activities, and other time-dependent loads. At the same time, building energy management must balance energy consumption, peak demand, and operational requirements, including occupant comfort where applicable [3,4].

Demand response (DR) provides a mechanism for making part of this building demand controllable. DR enables consumers to modify the timing or magnitude of electricity consumption in response to electricity prices, utility signals, or system conditions [6–8]. Optimization-based DR has been used to schedule flexible appliances and loads, reduce electricity payments, and reshape load profiles [7–9]. The controllability of electrical loads has also been studied from a broader power-system perspective, while specific load models have been developed to represent the response of HVAC and other controllable loads [10–13]. These studies establish the basic principle of using load flexibility as an energy-management resource rather than treating the building load as completely fixed.

The flexibility available from DR alone, however, is limited by the operating requirements of the loads. HVAC and lighting systems, for example, cannot be reduced or shifted without limits, and the available flexibility depends on equipment characteristics, operating schedules, and acceptable service conditions [10,11]. Once these limits are reached, additional load flexibility cannot be obtained simply by increasing the DR signal. Battery energy storage systems (BESSs) therefore provide a complementary source of flexibility. A BESS can charge during periods of low-cost electricity or available photovoltaic generation and discharge during high-cost or high-demand periods, providing energy shifting and peak-demand reduction [13–15].

The coordinated operation of flexible demand and energy storage has consequently become an established energy-management problem. Atzeni et al. [16] considered demand-side management together with distributed generation and storage, while Wang et al. [9] developed a coordinated price-based DR and BESS framework that explicitly considered battery state-of-charge management and uncertainty. Other building and microgrid studies have similarly combined renewable generation, storage, controllable loads, and grid transactions to reduce operating cost [17–20]. These studies demonstrate that DR and storage can provide complementary flexibility, but they also show that the resulting economic benefit depends on the building load profile, resource configuration, and electricity-market or tariff structure.

For commercial and institutional buildings, the electricity tariff is particularly important because the electricity bill may contain both volumetric energy charges and demand charges. A tariff with a demand component assigns a monetary cost to the customer's maximum demand over a specified billing period. Consequently, a short-duration reduction in the building's peak grid demand can have a much larger effect on the monthly bill than a comparable reduction in total energy consumption. Previous work has explicitly incorporated demand charges into BESS scheduling and building energy-management problems [20,21]. For example, Gao et al. formulated BESS scheduling as an MILP that includes energy cost, demand charge, and battery wear cost and evaluated the approach using measured campus-microgrid data [20].

The importance of the tariff structure has also been examined directly in commercial-building optimization. Hasan et al. developed a framework for optimizing commercial-building loads with renewable generation and BESS under different utility rate structures and showed that the tariff type can strongly influence the economic value of storage [22]. More recent work has extended this idea to multiple tariff categories and analyzed how changes in energy and demand-charge components affect optimization results [22,23]. These studies make clear that the value of DR and BESS cannot be separated from the billing mechanism under which a facility operates. However, tariff-oriented studies generally consider generalized tariff classes or collections of commercial-building profiles, whereas a facility-specific evaluation using the actual utility tariff and measured operating data provides a different practical perspective.

This issue is especially relevant for university campuses, where buildings combine laboratories, classrooms, offices, HVAC systems, lighting, and other loads with different degrees of flexibility. Existing campus studies have investigated price-based DR, photovoltaic generation, storage, and energy-management systems using optimization-based approaches [18,19,24]. Real-building studies have also demonstrated the value of measured smart-meter data for building energy-management optimization [19]. In addition, commercial-building studies have shown that peak demand and load-profile characteristics can strongly affect the economic value of BESS deployment [15,20,21]. These results motivate a case-specific assessment in which the optimization is tied directly to the measured load of an institutional building and the tariff actually applied to that facility.

Motivated by this practical need, this paper develops an optimal energy-management framework for the Engineering Laboratory Building (ELB) at the University of Michigan-Dearborn. The building is supplied by the utility grid and an existing rooftop photovoltaic system, while a BESS is modeled as a proposed additional resource. The energy-management problem is formulated as a mixed-integer linear programming (MILP) model that jointly schedules flexible building demand and BESS charging/discharging subject to demand-response, battery, PV, grid-capacity, and power-balance constraints. The optimization is performed over the monthly billing period so that the demand-charge component can be represented directly. The electricity cost is calculated using the DTE Energy Primary Educational Institution Rate D6.2, including its energy, demand, and fixed service charges [25].

The framework is evaluated using measured ELB electricity-demand data and measured rooftop-PV generation data. Four operating cases are compared under the same billing-period data and tariff: baseline operation, DR only, BESS only, and coordinated DR and BESS. This design isolates the contribution of demand flexibility, the contribution of storage, and the additional benefit obtained when the two resources are operated together. For the May 2026 case, DR alone reduces the monthly electricity bill by 12.0%, while coordinated DR and BESS operation reduces the bill by 25.8%. Additional BESS-capacity cases and a December 2025 case are used to examine how the observed savings vary with storage capacity and load-profile characteristics.

The main contributions of this work are summarized as follows:

1. A coordinated DR–BESS energy-management framework is formulated as a MILP problem for an institutional building, jointly scheduling flexible loads and BESS operation under practical operating constraints.
2. The actual commercial electricity tariff applied to the ELB is incorporated into the cost model, including its energy and demand-related billing components, so that peak-demand reduction is directly reflected in the optimization objective [22,25].
3. The framework is evaluated using measured electricity-demand and rooftop-PV data from a real campus building rather than relying exclusively on synthetic or benchmark profiles.
4. Baseline, DR-only, BESS-only, and coordinated DR–BESS cases are compared under the same building data and tariff to quantify the separate and combined economic contributions of load flexibility and storage.

## 2. System Configuration and Operation Strategy

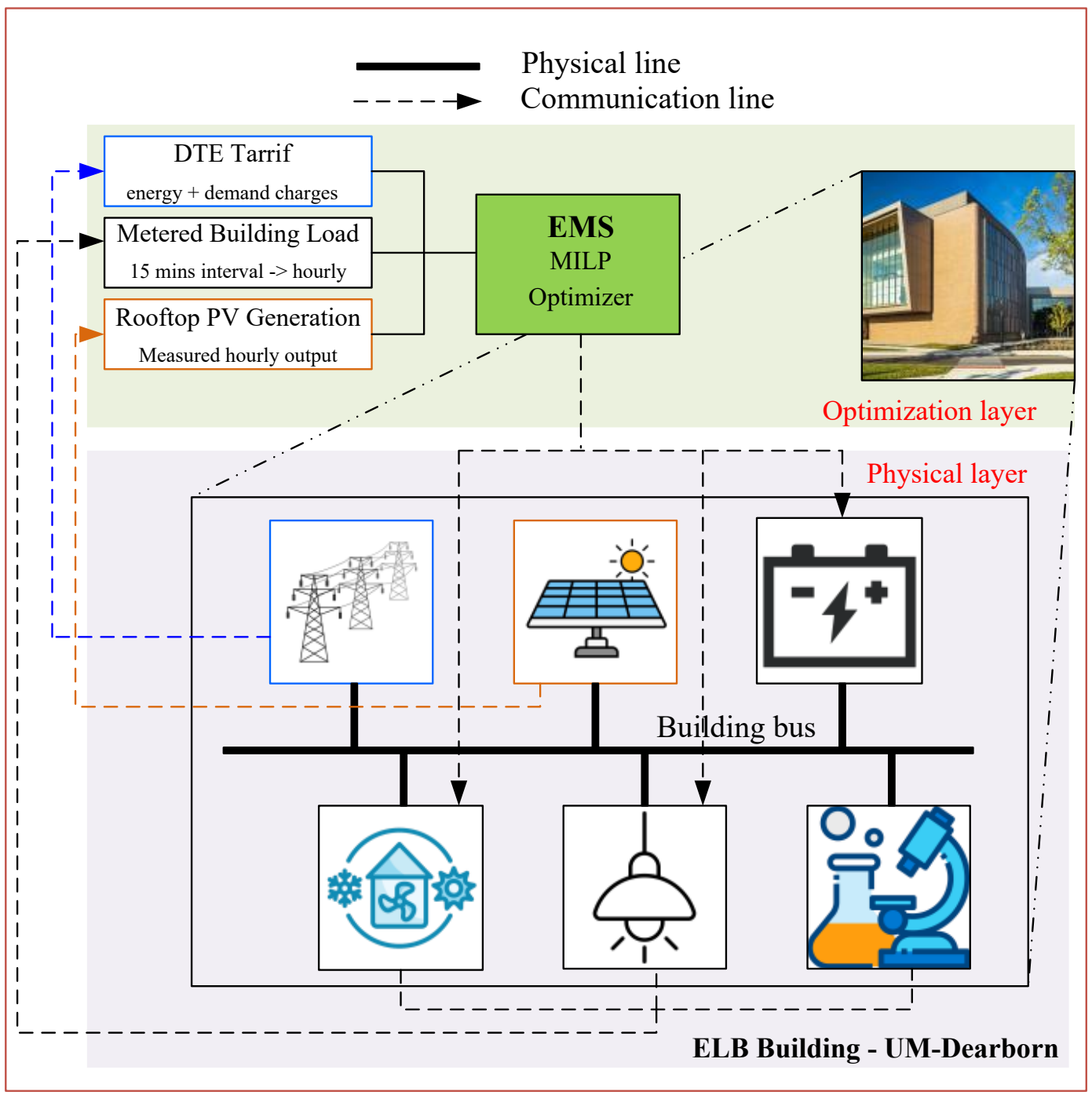

**Figure 1.** Building System Configuration.

Figure 1 illustrates the system configuration and operation strategy. The framework is split into two layers: the optimization layer and the physical layer. The upper layer contains the data and optimization components: measured building loads recorded at 15-minute intervals and aggregated to hourly intervals, measured hourly rooftop PV generation, and the DTE 6.2 Tariff. These inputs feed a mixed-integer linear program implemented in Python, which determines the hourly operating schedule for the billing period. The lower layer represents the physical building energy system. The ELB is supplied from the utility grid and from the existing rooftop solar array, both of which connect to the building bus. The battery energy storage system is modeled as a proposed addition rather than existing equipment and supplies power to the bus.

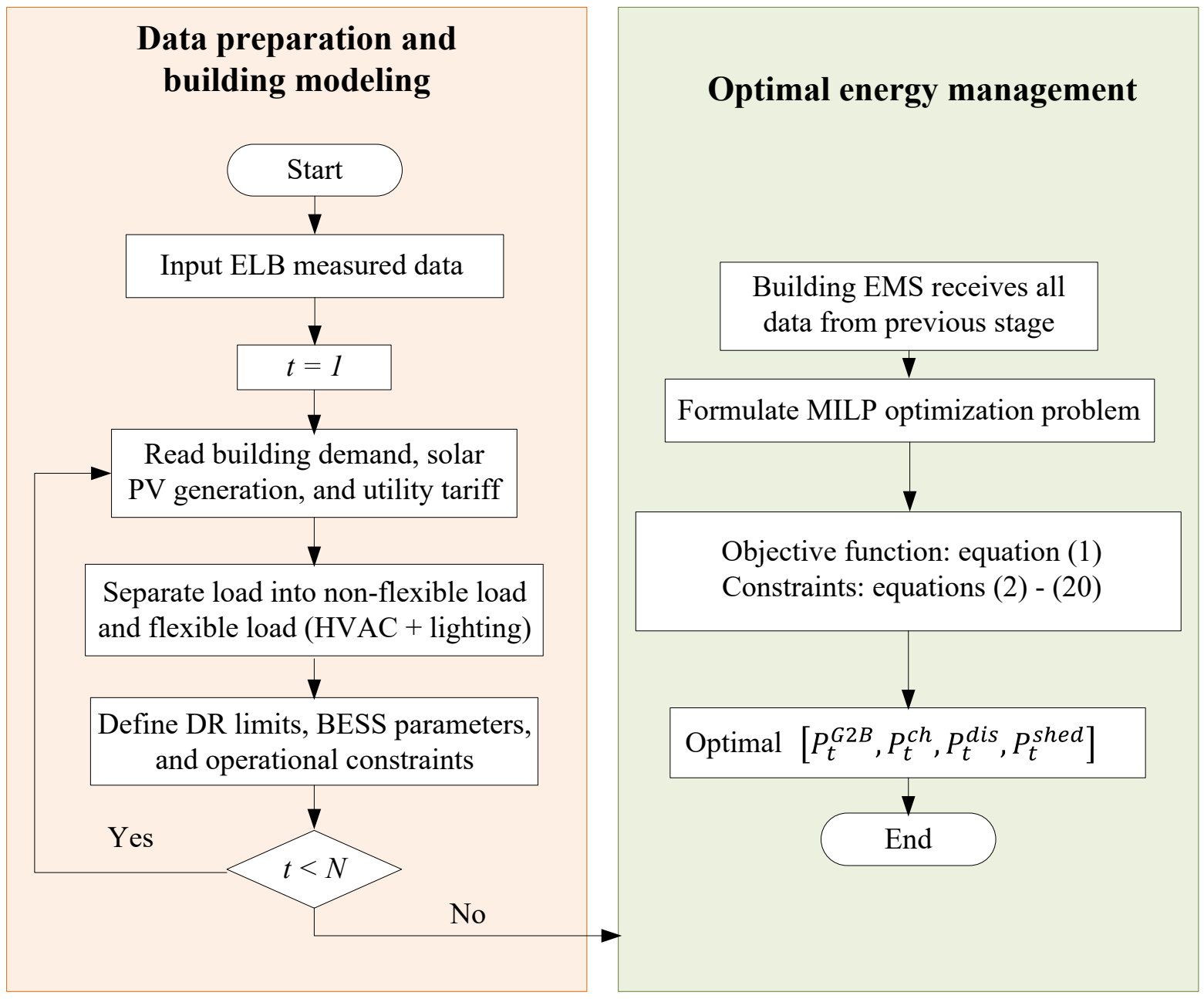

**Figure 2.** Flowchart for optimal operation of the campus building.

Figure 2 shows the overall data preparation and optimization procedure. The process is organized into two stages. In the first stage, we input measured load, solar generation, and utility tariff data for the ELB for each hour t of the billing month. At each hour, the building load is separated into a fixed component and a flexible component, where the flexible component consists of the HVAC and lighting loads available to demand response. Demand response limits, BESS parameters, and other operational constraints are then defined. This process repeats for every hour until t = N, after which the fully prepared dataset is passed to the optimization stage.

In the second stage, the building energy management system receives the complete dataset and formulates the MILP optimization problem described in Section 3. The objective minimizes total monthly electricity cost, comprising the energy charge, demand charge, and fixed service charge, subject to constraints on demand response, PV utilization, BESS operation, and power balance. The optimizer jointly determines the flexible load schedule and the BESS charging and discharging schedule, from which the optimal grid import, PV utilization, and monthly billed peak demand are obtained. The final outputs are the optimal flexible load schedule, BESS charge and discharge profile, grid import profile, PV utilization, peak demand, and total monthly electricity cost.

## 3. Problem Formulation

The proposed framework is formulated as a mixed-integer linear program (MILP) that determines the optimal ELB schedule over a monthly billing cycle. The building is currently supplied by grid electricity and on-site solar PV generation, and a portion of the load is treated as flexible and available to demand response. Additionally, we propose a battery energy storage system (BESS) as an additional resource to assist demand response. A billing cycle is defined by a set of hourly intervals T = {0,1,…, N-1} where N is the number of hours in a billing month. We used a monthly cycle rather than a daily cycle because the demand charge tariff is based on the single highest demand within the billing period and cannot be captured in a daily window.

The objective is to minimize the total operational cost of the building. In Equation (1), starting from the left, the first term is the energy charge for power imported from the grid into the system. The second and third terms are the charging and discharging degradation costs of the BESS. The final terms are the demand charge that is applied to the monthly peak and the fixed monthly service charge, both of which are determined by the tariff rather than by hourly consumption. Load shed by demand response does not appear as a separate term. It reduces the power imported from the grid through the power balance in Equation (20), so its savings are already captured in the energy charge and, during on-peak hours, in the demand charge.

$$\sum_{t=1}^{T}\left(p_t^{G2B}.c_t^{grid} + c_c.p_t^{ch} + c_{dc}.p_t^{dis}\right) + \left(c^{cap} + c^{noncap} + c^{dist}\right)P^{peak} + c^{svc} \quad (1)$$

where:
$p_t^{G2B}$ is the power imported from the grid

$c_t^{grid}$ is the grid energy price
$c_c, c_{dc}$ are the charging and discharging degradation costs of the BESS
$p_t^{ch}, p_t^{dis}$ are the charging and discharging power of the BESS
$P^{peak}$ is the monthly billed peak demand
$c^{cap}, c^{noncap}, c^{dist}$ are the capacity, non-capacity, and distribution demand charge rates
$c^{svc}$ is the fixed monthly service charge

Demand response is represented by two mechanisms: load shedding, in which demand is reduced and not recovered, and load shifting, where demand is relocated to a lower peak hour. The load shed each hour is limited by the flexibility of the load, as given by (2). The maximum sheddable load each hour is the sum of each hour's flexible share, as expressed in (3), where the HVAC and lighting load profiles are defined by (4) and (5). We also limit the shedding to a predetermined demand response event window by (6), limiting the shedding to the peak hours.

Load shifting is modeled using two variables representing demand removed and added to each hour. The quantity shifted out is limited by the flexibility of the system, and the quantity shifted in has additional headroom so that the specified hour can take in the shifted demand without overloading, as expressed in (7) and (8). Because shifting relocates demand in time rather than eliminating it, the total shifted out over the cycle must be equal to the amount shifted in; this is enforced by (9).

$$p_t^{shed} \leq S_t^{max} \quad \forall\, t \in T \tag{2}$$

$$S_t^{max} = \alpha^{hvac} L_t^{hvac} + \alpha^{light} L_t^{light} \quad \forall\, t \in T \tag{3}$$

$$L_t^{hvac} = L_t \cdot \mu_t^{hvac} \quad \forall\, t \in T \tag{4}$$

$$L_t^{light} = L_t \cdot \mu_t^{light} \quad \forall\, t \in T \tag{5}$$

$$S_t^{max} = 0 \quad \forall\, t \notin [t_{start}, t_{end}] \tag{6}$$

$$p_t^{out} \leq F_t^{max} \quad \forall\, t \in T \tag{7}$$

$$p_t^{in} \leq F_t^{max} + 0.25\, L_t \quad \forall\, t \in T \tag{8}$$

$$\sum_{t=1}^{T} p_t^{out} = \sum_{t=1}^{T} p_t^{in} \tag{9}$$

where:
$S_t^{max}$ is the maximum sheddable load at t
$L_t$ is the measured building load at t
$p_t^{shed}$ is the amount of load shed due to demand response
$L_t^{hvac}, L_t^{light}$ are the HVAC and lighting load at t
$\alpha^{hvac}, \alpha^{light}$ are the allowed shed fractions of the HVAC and lighting systems
$t_{start}, t_{end}$ are the start and end of the demand response event window
$p_t^{out}, p_t^{in}$ are the amounts of load shifted out of and into t
$F_t^{max}$ is the flexible load available for shifting at t

The PV power utilized by the building cannot exceed the amount of solar generated for any given hour, as represented by Equation (10). We use equations (11)-(17) to constrain our BESS. In (11) and (12), charging and discharging power are capped and only permitted when they are active. Equation (13) prevents a situation where we are charging and discharging the BESS at the same time. The stored energy carries over from hour to hour in Equations (14) and (15), with the first interval initialized from the starting charge. Equation (16) requires the battery to end the billing cycle at the level it began so that the schedule is self-contained, and Equation (17) holds the stored energy within the physical limits of the battery.

$$p_t^{PV} \leq PV_t \quad \forall\, t \in T \tag{10}$$

$$p_t^{ch} \leq I_t^{ch} P^{ch,max} \quad \forall\, t \in T \tag{11}$$

$$p_t^{dis} \leq I_t^{dis} P^{dis,max} \quad \forall\, t \in T \tag{12}$$

$$I_t^{ch} + I_t^{dis} \leq 1 \quad \forall t \in T \tag{13}$$

$$E_0 = E^{init} + \eta^{ch} p_0^{ch} - \frac{p_0^{dis}}{\eta^{dis}} \tag{14}$$

$$E_t = E_{t-1} + \eta^{ch} p_t^{ch} - \frac{p_t^{dis}}{\eta^{dis}} \quad \forall t \geq 1 \tag{15}$$

$$E_0 = E_{|T|-1} \tag{16}$$

$$E^{min} \leq E_t \leq E^{max} \quad \forall t \in T \tag{17}$$

where:
$p_t^{PV}$ is the PV power utilized by the building at t
$PV_t$ is the available solar generation at t
$I_t^{ch}, I_t^{dis}$ are the binary charging and discharging indicators at t
$P^{ch,max}, P^{dis,max}$ are the maximum charging and discharging power of the BESS
$E_t$ is the energy stored in the BESS at t
$E^{init}$ is the initial stored energy of the BESS
$\eta^{ch}, \eta^{dis}$ are the charging and discharging efficiencies
$E^{min}, E^{max}$ are the minimum and maximum stored energy of the BESS

Grid import is bounded by the service capacity of the building, as given by Equation (18). Then Equation (19) defines the billed peak demand, which is required to be as large as the grid import in every on-peak hour. Since Equation (1) is applied to these values, the optimization function uses this as a common ceiling for everything to follow under. The operation of the building has to follow the power balance rule; for each interval, the amount we input, generate through PV, and discharge from the BESS must be equal to the building load plus the BESS charging, as given by equation (20)

$$p_t^{G2B} \leq P^{G2B,max} \quad \forall t \in T \tag{18}$$

$$P^{peak} \geq p_t^{G2B} \quad \forall t \in T_{on} \tag{19}$$

$$p_t^{G2B} + p_t^{PV} + \eta^{dis} p_t^{dis} = \left(L_t - p_t^{shed} - p_t^{out} + p_t^{in}\right) + \frac{p_t^{ch}}{\eta^{ch}} \quad \forall t \in T \tag{20}$$

where:
$P^{G2B,max}$ is the maximum grid connection capacity of the building
$T_{on}$ is the set of on-peak hours defined by the tariff

## 4. Engineering Laboratory Building (ELB) and Dataset

This framework is validated by using the Engineering Laboratory Building (ELB) at the University of Michigan-Dearborn as a real-world case study. The ELB is a 123,000 square-foot facility including a mix of engineering laboratories and classroom space. The building is billed under the DTE tariff structure, and the building's load is monitored through dedicated electrical metering, which forms the basis of the dataset used in this study.

The data provided was recorded at 15-minute intervals over a multi-year period, with no missing values, leading to consistent data. Consistent with the hourly resolution of the optimization framework, the 15-minute readings were aggregated to hourly average power values. The solar PV generation was also recorded separately at hourly readings beginning in September 2025. For this study, we selected two months for testing: May 2026, representing the higher-demand cooling season, and December 2025, representing the lower-demand winter season. Each month provides 744 hourly load and solar generation data points, and both fall within the period covered by the PV data.

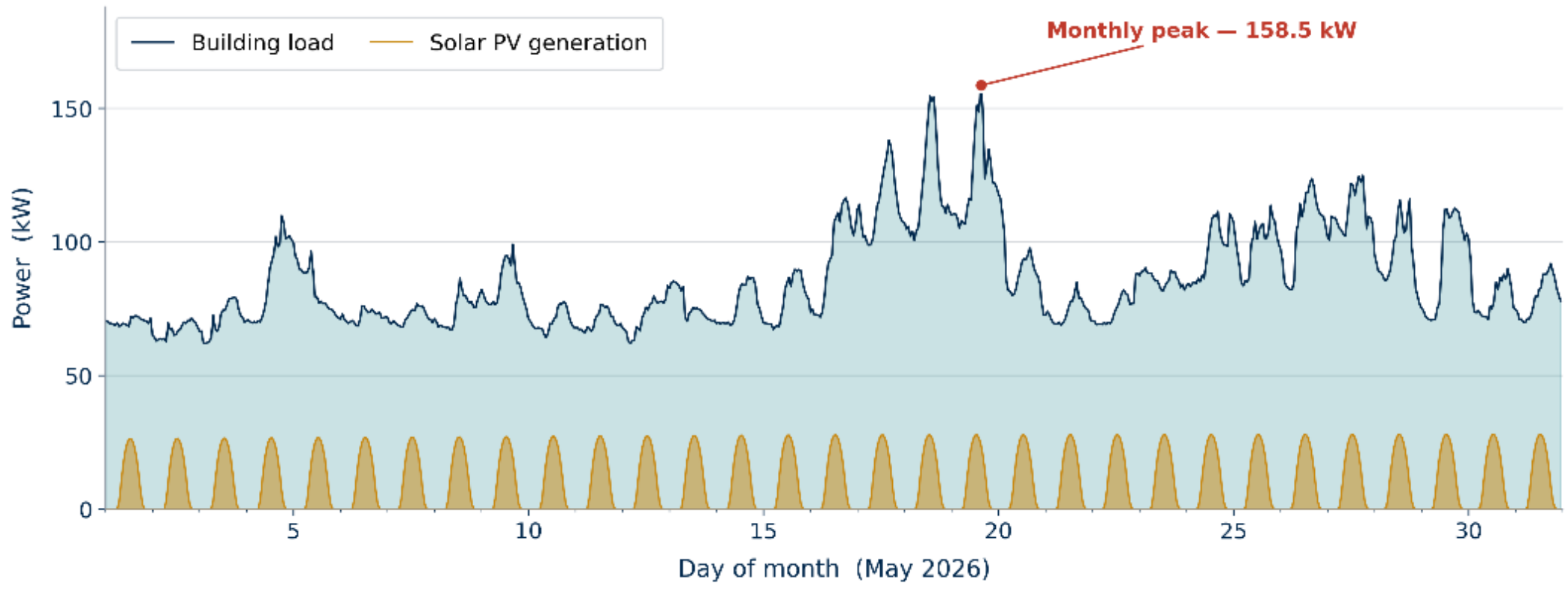

**Figure 3.** The monthly load distribution of the ELB for the month of May.

**Table 1.** Load of the ELB comparing weekends to weekdays in May.

| | May Weekday | May Weekend |
|---|---|---|
| Days | 21 | 10 |
| Average Load | 86.6kW | 84.3kW |
| Maximum Load | 158.5kW | 138.0kW |

Based on the data provided, we can see that the ELB maintains a large, continuous operating base load; this is consistent with laboratory equipment, HVAC, and lighting being active even on the weekend, as they can be serving research spaces.

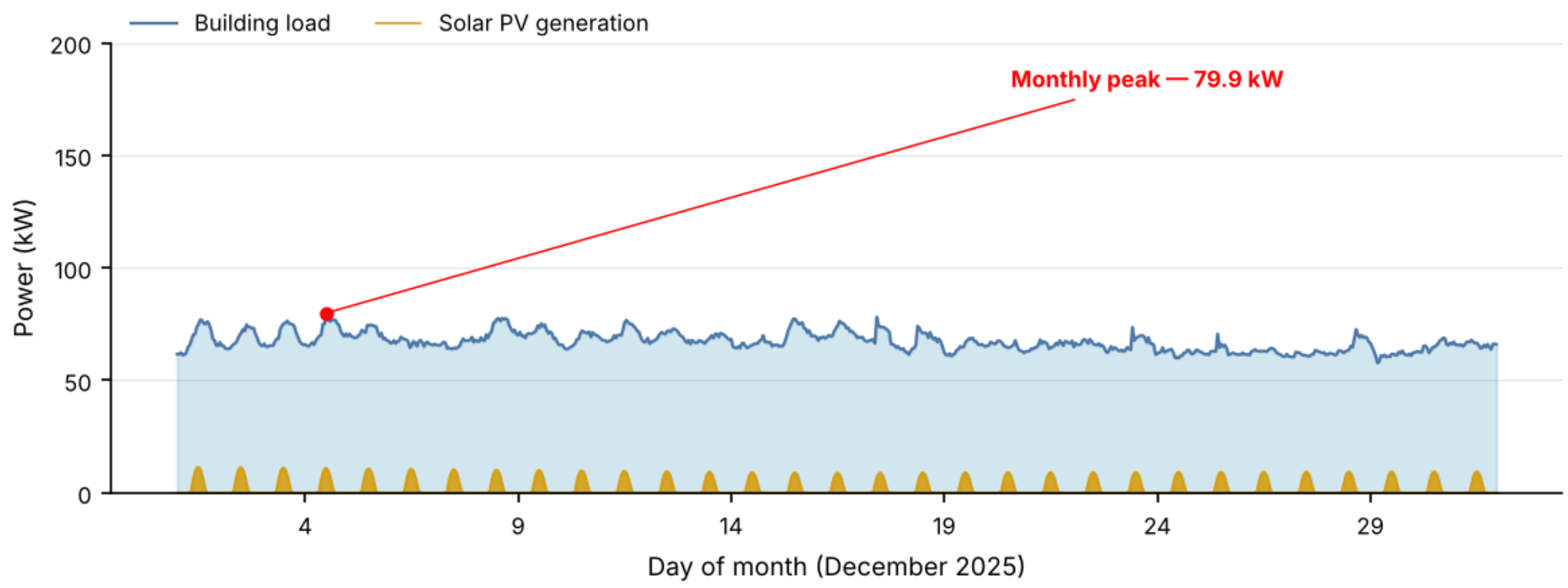

**Figure 4.** The monthly load distribution of the ELB for the month of December.

**Table 2.** Load of the ELB comparing weekends to weekdays in December.

| | December Weekday | December Weekend |
|---|---|---|
| Days | 23 | 8 |
| Average Load | 67.7kW | 65.6kW |
| Maximum Load | 79.9kW | 72.6kW |

Similar to Table 1, the weekend average was very similar; for the weekday, the average load was 67.7kW, while the weekend was 65.6kW.

The ELB is billed under DTE Energy's Primary Educational Institution Rate (D6.2) [25], a commercial tariff that combines energy time charges with a demand charge based on the customer's peak power draw during on-peak hours. The complete rate structure, obtained from the building's March 2026 electricity invoice, is summarized in the table below.

**Table 3.** DTE energy tariff D6.2.

| Parameters | Rate |
|---|---|
| Energy (On-Peak) | \$0.04535/kWh |
| Energy (Off-Peak) | \$0.04235/kWh |
| Capacity Demand Charge | \$10.34/kW |
| Non-Capacity Demand Charge | \$3.99/kW |

| Parameters | Rate |
|---|---|
| Distribution Demand Charge | $7.15/kW |
| Service Charge | $70.00/month |

In the above table the energy charge is billed per kilowatt-hour consumed and differs only slightly between on-peak and off-peak hours ($0.04535/kWh and $0.04235/kWh, respectively), a spread of about 7%. Because the difference is small compared to typical time-of-use rate structures seen in residential tariffs, shifting energy consumption between hours provides limited direct benefit under this tariff.

The demand charges, however, are billed on the single highest power demand recorded during on-peak hours within the billing month, independent of total energy consumed. Three components apply: a capacity demand charge of $10.34/kW, a non-capacity demand charge of $3.99/kW, and a distribution demand charge of $7.15/kW, together totaling $21.48/kW of monthly billed peak demand. A fixed monthly service charge of $70.00 is also applied regardless of consumption or demand.

Because the demand charges are assessed on a single hour of the month while the energy charge accumulates gradually over 744 hours, a short-duration reduction in peak demand can produce a cost benefit comparable to a much larger reduction in total energy consumption. This asymmetry motivates the demand-charge-aware objective-function formulation presented in Section 3 and is shown in Section 5 to account for most of the cost savings achieved in this study.

## 5. Simulation Results and Discussion

A mixed-integer linear program (MILP) was developed in Python using the PuLP modeling library and solved with a CBC solver. All of our cases were evaluated over two 744-hour billing periods, May 2026 and December 2025, using the measured building load, solar generation data, and the DTE D6.2 tariff parameters for each month. For our proposed BESS, we modeled a 200 kWh usable capacity as the primary case, with additional 100 kWh and 150 kWh cases used to examine how the savings change with storage capacity. We compared four different usage cases. The baseline case represented our normal operation in which no demand response or storage is applied. We obtained this by setting our shedding and shifting ceilings to zero. The next case was utilizing demand response only; this enables load shedding and shifting but provides no battery storage. The BESS-only case enables battery scheduling while following scenario one and disabling shifting and shedding. Lastly, the DR + BESS case enables both resources simultaneously.

Table 4 summarizes the parameters used across all four cases. We simulated with a BESS of 200kWh of usable capacity and a minimum reserve of 20kWh to prevent full discharge. The charging and discharging efficiencies were taken to be 92%, which was around 85% for a full cycle. In order to keep the cycle self-contained, the battery's initial charge was set to half of the rated capacity, and at the end of the billing period it was set to return to this level. For the demand response cases, HVAC, which is often one of the largest consumers of power, was assumed to represent 45% of the total building load, and lighting was assumed to be around 35 kW, about 22% of peak demand. HVAC was permitted to be reduced by up to 30% and lighting by up to 15% during the demand response event window.

**Table 4.** Model Parameters.

| Parameters | Value |
|---|---|
| BESS Capacity | 200kWh |
| BESS Minimum | 10% of capacity (20 kWh at 200 kWh) |
| Initial charge | 100kWh (50% of capacity) |
| HVAC Load | 45% |
| Maximum HVAC reduction | 30% |
| Lighting Load | 35kW |
| Maximum Lighting reduction | 15% |
| Maximum grid import | 250kW |

The table below summarizes the monthly electricity cost of each case. Looking at the baseline, we can see that the total bill was $5471, with the demand charge being around $2889, which is 53% of the bill despite only being a single hour of operation. This confirms that peak management, rather than conserving energy, is the dominant cost saver.

**Table 5.** Monthly bill comparison for May across DR, BESS, and combined scenarios (200 kWh battery).

| Scenario | Bill | Energy Charge | Demand Charge | Billed Peak | Savings (%) |
|---|---|---|---|---|---|
| Baseline | $5,471.47 | $2,512.84 | $2,888.63 | 134.5kW | — |
| DR Only | $4,815.83 | $2,340.81 | $2,405.02 | 112.0kW | 12.0 |
| BESS Only | $4,942.91 | $2,515.22 | $2,357.69 | 109.8kW | 9.7 |
| DR + BESS | $4,062.41 | $2,351.28 | $1,641.13 | 76.4kW | 25.8 |

Demand response alone reduces the monthly bill to $4,816, a 12.0% saving relative to the baseline. This saving comes from both billing components: shedding removes energy from the billing period, lowering the energy charge, while shedding and shifting reduce the billed peak from 134.5 kW to 112.0 kW and lower the demand charge by $483. The BESS-only case behaves differently; however, while the peak rate was lower overall, we see that the bill was still higher than using DR alone. This can be assumed to be because a battery offsets our peak load, but we are still requiring the same amount of energy without cutting down. In our last scenario, coordinated operation of both resources produces the largest saving, reducing the monthly bill to $4,062, or 25.8% below the baseline. The improvement over either resource individually is greater than a simple sum of their separate contributions would suggest: DR alone and BESS alone save 12.0% and 9.7%, respectively, and the two together save 25.8%. Having both systems complement each other rather than being redundant, Demand Response reduces the peak load by shifting and shedding, and the battery then discharges against the reduced profile that remains, so each kilowatt of battery discharge is applied to a lower starting peak.

**Table 6.** Monthly bill comparison for May across a baseline, BESS, and combined scenarios (100 kWh battery).

| Scenario | Bill | Energy Charge | Demand Charge | Billed Peak | Savings (%) |
|---|---|---|---|---|---|
| Baseline | $5,471.47 | $2,512.84 | $2,888.63 | 134.5kW | — |
| BESS Only | $5,098.14 | $2,513.78 | $2,514.36 | 117.1 kW | 6.8% |
| DR + BESS | $4,374.70 | $2,342.16 | $1,962.54 | 91.4 kW | 20.0% |

Table 6 presents the same tests but for a 100 kWh battery. With storage alone, we see that the bill falls to $5,098.14, a 6.8% saving, while the peak is reduced from 134.5 kW to 117.1 kW. When combined with demand response, the bill falls to $4,374.70, corresponding to a 20.0% saving and a billed peak of 91.4 kW. Similar to the 200 kWh case, the energy charge in the BESS-only scenario remains essentially unchanged from the baseline, $2,513.78 against $2,512.84, confirming that storage relocates energy in time rather than reducing total consumption. All of the savings in that case are therefore due to the reduction in demand charge.

**Table 7.** Monthly bill comparison for May across a baseline, BESS, and combined scenarios (150 kWh battery).

| Scenario | Bill | Energy Charge | Demand Charge | Billed Peak | Savings (%) |
|---|---|---|---|---|---|
| Baseline | $5,471.47 | $2,512.84 | $2,888.63 | 134.5kW | — |
| BESS Only | $5,018.40 | $2,514.36 | $2,434.04 | 113.3 kW | 8.3% |
| DR + BESS | $4,214.78 | $2,345.87 | $1,798.91 | 83.7 kW | 23.0% |

Table 7 presents the results for a 150 kWh battery. Storage alone reduces the bill to $5,023.89, an 8.2% saving with a billed peak of 113.3 kW, while combined operation reduces the bill to $4,077.87, a 25.5% saving with a billed peak of 83.7 kW.

Comparing Tables 5 through 7 shows two consistent trends. First, increasing battery capacity produces a reduction in the peak and the total cost; however, it is not as effective as demand response. In all three battery scenario tests, the cost savings came from the demand charge since we were shifting the energy to a different point in time, which, while effective, wasn't the best solution. Even at 200kWh, the BESS-only case gets a 9.7% saving compared to demand response reaching 12.0%. This supports the conclusion that demand response should be the primary cost-saving measure, with storage deployed as a complementary resource.

**Table 8.** Monthly bill comparison for December across DR, BESS, and combined scenarios (200 kWh battery).

| Scenario | Bill | Energy Charge | Demand Charge | Billed Peak | Savings (%) |
|---|---|---|---|---|---|
| Baseline | $3,881.81 | $2,147.11 | $1,664.70 | 77.5 kW | — |

| Scenario | Bill | Energy Charge | Demand Charge | Billed Peak | Savings (%) |
|---|---|---|---|---|---|
| DR Only | $3,436.09 | $2,006.80 | $1,359.30 | 63.3 kW | 11.5 |
| BESS Only | $3,569.66 | $2,174.44 | $1,325.22 | 61.7 kW | 8.0 |
| DR + BESS | $3,028.82 | $2,051.40 | $907.42 | 42.2 kW | 22.0 |

Demand response produced nearly identical relative savings in both months, 12.0% in May and 11.5% in December; this shows that the benefit of load shedding and shifting is not season-dependent and can occur year-round. The value of battery storage, however, was lower in winter: the combined DR and BESS strategy achieved a 22.0% reduction in December compared with 25.8% in May. This difference arises because the December load profile is considerably flatter than the May profile, with a baseline peak of only 77.5 kW compared with 134.5 kW, leaving less peak demand available for the battery to shave once demand response has already reduced the load profile.

**Table 9.** Monthly bill comparison for December across a baseline, BESS, and combined scenarios (100 kWh battery).

| Scenario | Bill | Energy Charge | Demand Charge | Billed Peak | Savings (%) |
|---|---|---|---|---|---|
| Baseline | $3,881.81 | $2,147.11 | $1,664.70 | 77.5kW | — |
| BESS Only | $3,653.79 | $2,156.09 | $1,427.70 | 66.5kW | 5.9% |
| DR + BESS | $3,119.80 | $2,032.33 | $1,017.47 | 47.4kW | 19.6% |

With storage alone at 100 kWh, the bill falls to $3,653.79, a 5.9% saving, with a billed peak reduced from 77.5 kW to 66.5 kW. Combined with demand response, the bill falls to $3,119.80, a 19.6% saving, with a billed peak of 47.4 kW. As in May, the BESS-only energy charge remains close to the baseline value ($2,156.09 versus $2,147.11), confirming that the savings in that case are driven almost entirely by the reduced demand charge rather than reduced energy consumption.

**Table 10.** Monthly bill comparison for December across a baseline, BESS, and combined scenarios (150 kWh battery).

| Scenario | Bill | Energy Charge | Demand Charge | Billed Peak | Savings (%) |
|---|---|---|---|---|---|
| Baseline | $3,881.81 | $2,147.11 | $1,664.70 | 77.5kW | — |
| BESS Only | $3,610.40 | $2,164.05 | $1,376.35 | 64.1kW | 7.0% |
| DR + BESS | $3,073.33 | $2,041.26 | $962.07 | 44.8kW | 20.8% |

At 150 kWh, storage alone yields a 7.0% saving with a billed peak of 64.1 kW, while coordinated operation reaches 20.8% with a billed peak of 44.8 kW. Comparing Tables 8 through 10 with their May counterparts (Tables 5 through 7) shows the same qualitative pattern in both months: diminishing marginal benefit with increasing capacity, and demand response alone outperforming storage alone at every capacity tested, confirming that the underlying mechanism is not specific to a single month

**Table 11.** Comparison of bill, peak demand, and savings for May and December under DR+BESS at increasing battery capacity.

| Capacity (kWh) | May 2026 | | | December 2025 | | |
|---|---|---|---|---|---|---|
| | Bill ($) | Peak (kW) | Savings (%) | Bill ($) | Peak (kW) | Savings (%) |
| DR Only | 4,815.83 | 112.0 | 12.0 | 3,436.09 | 63.3 | 11.5 |
| 100 | 4,374.70 | 91.4 | 20.0 | 3,119.80 | 47.4 | 19.6 |
| 150 | 4,214.78 | 83.7 | 23.0 | 3,073.33 | 44.8 | 20.8 |
| 200 | 4,062.41 | 76.4 | 25.8 | 3,028.82 | 42.2 | 22.0 |

At every capacity, both the bill and the billed peak are consistently lower in December than in May, reflecting December's smaller baseline load. However, the percentage savings diverge as capacity increases; the two months differ by only 0.5 points at DR-only (12.0% vs. 11.5%) but diverge to 3.8 points at 200 kWh (25.8% vs. 22.0%), since the flatter December load profile leaves less peak available for the battery to shave once demand response has already been applied.

## 6. Conclusions

This paper presented a MILP-based energy management framework for campus buildings that schedules demand response and battery energy storage under an accurate commercial electricity tariff. Through a study of the Engineering Lab Building, we concluded that peak power management, rather than total energy consumption, is the primary driver of cost for institutional buildings billed with a demand charge. Demand response alone reduced the monthly bill by 12.0%, while the coordinated operation of demand response and battery storage achieved a 25.8% reduction and lowered the billed peak from 134.5 kW to 76.4 kW. Across the 4 cases presented, the energy charge varied by less than 7%, while the demand charge fell by $1,248; this confirmed that the savings originate almost entirely from the demand charge of the bill. However, it is also important to note that demand response alone outperformed the battery alone, 12.0% against 9.7%, because shed load is removed entirely; it offsets both the peak and the energy component, compared to storage, which just relocates it to another point in time. But the two were not redundant; rather, they were complementary. As mentioned above, their combined operation achieved a 25.8% reduction as demand response lowers the load profile against which the battery then discharges. Based on the results, this suggests that demand response, which requires no investment, represents the logical first step for institutional buildings, with storage deployed as a complement to the existing system. Future work will extend our research into a full-year evaluation and determine the optimal battery and payback period.

**Acknowledgements**

This work was supported by the Summer Undergraduate Research Experience (SURE) Program, University of Michigan-Dearborn.

## References

1. International Energy Agency. Buildings. In *Energy Efficiency 2025*; IEA: Paris, France, 2025.
2. Zia, M.F.; Elbouchikhi, E.; Benbouzid, M.E.H. Microgrids energy management systems: A critical review on methods, solutions, and prospects. *Appl. Energy* **2018**, *222*, 1033–1055. https://doi.org/10.1016/j.apenergy.2018.04.103.
3. Oldewurtel, F.; Parisio, A.; Jones, C.N.; Gyalistras, D.; Gwerder, M.; Stauch, V.; Lehmann, B.; Morari, M. Use of model predictive control and weather forecasts for energy efficient building climate control. *Energy Build.* **2012**, *45*, 15–27. https://doi.org/10.1016/j.enbuild.2011.09.022.
4. Killian, M.; Kozek, M. Ten questions concerning model predictive control for energy efficient buildings. *Build. Environ.* **2016**, *105*, 403–412. https://doi.org/10.1016/j.buildenv.2016.05.034.
5. Drgoňa, J.; et al. All you need to know about model predictive control for buildings. *Annu. Rev. Control* **2020**, *50*, 190–232. https://doi.org/10.1016/j.arcontrol.2020.09.001.
6. Palensky, P.; Dietrich, D. Demand side management: Demand response, intelligent energy systems, and smart loads. *IEEE Trans. Ind. Inform.* **2011**, *7*, 381–388. https://doi.org/10.1109/TII.2011.2158841.
7. Mohsenian-Rad, A.-H.; Wong, V.W.S.; Jatskevich, J.; Schober, R.; Leon-Garcia, A. Autonomous demand-side management based on game-theoretic energy consumption scheduling for the future smart grid. *IEEE Trans. Smart Grid* **2010**, *1*, 320–331. https://doi.org/10.1109/TSG.2010.2089069.
8. Pedrasa, M.A.A.; Spooner, T.D.; MacGill, I.F. Coordinated scheduling of residential distributed energy resources to optimize smart home energy services. *IEEE Trans. Smart Grid* **2010**, *1*, 134–143. https://doi.org/10.1109/TSG.2010.2053053.
9. Wang, B.; Zhang, C.; Dong, Z.Y. Interval optimization based coordination of demand response and battery energy storage system considering SOC management in a microgrid. *IEEE Trans. Sustain. Energy* **2020**, *11*, 2922–2931. https://doi.org/10.1109/TSTE.2020.2982205.
10. Zhang, W.; Lian, J.; Chang, C.-Y.; Kalsi, K. Aggregated modeling and control of air conditioning loads for demand response. *IEEE Trans. Power Syst.* **2013**, *28*, 4655–4664. https://doi.org/10.1109/TPWRS.2013.2266121.
11. Shao, S.; Pipattanasomporn, M.; Rahman, S. Development of physical-based demand response-enabled residential load models. *IEEE Trans. Power Syst.* **2013**, *28*, 607–614. https://doi.org/10.1109/TPWRS.2012.2208232.
12. Callaway, D.S.; Hiskens, I.A. Achieving controllability of electric loads. *Proc. IEEE* **2011**, *99*, 184–199. https://doi.org/10.1109/JPROC.2010.2081652.
13. Sæle, H.; Grande, O.S. Demand response from household customers: Experiences from a pilot study in Norway. *IEEE Trans. Smart Grid* **2011**, *2*, 102–109. https://doi.org/10.1109/TSG.2010.2104165.
14. U.S. Department of Energy. *On-Site Energy Storage Decision Guide*; Better Buildings Initiative, U.S. Department of Energy: Washington, DC, USA, 2017.
15. Leadbetter, J.; Swan, L. Battery storage system for residential electricity peak demand shaving. *Energy Build.* **2012**, *55*, 685–692. https://doi.org/10.1016/j.enbuild.2012.09.035.
16. Atzeni, I.; Ordóñez, L.G.; Scutari, G.; Palomar, D.P.; Fonollosa, J.R. Demand-side management via distributed energy generation and storage optimization. *IEEE Trans. Smart Grid* **2013**, *4*, 866–876. https://doi.org/10.1109/TSG.2012.2206060.
17. Yan, Q.; Zhang, B.; Kezunovic, M. Optimized operational cost reduction for an EV charging station integrated with battery energy storage and PV generation. *IEEE Trans. Smart Grid* **2019**, *10*, 2096–2106. https://doi.org/10.1109/TSG.2017.2788440.
18. Muqeet, H.A.U.; Ahmad, A. Optimal scheduling for campus prosumer microgrid considering price based demand response. *IEEE Access* **2020**, *8*, 71378–71394. https://doi.org/10.1109/ACCESS.2020.2987915.
19. Thomas, D.; Deblecker, O.; Ioakimidis, C.S. Optimal operation of an energy management system for a grid-connected smart building considering photovoltaics' uncertainty and stochastic electric vehicles' driving schedule. *Appl. Energy* **2018**, *210*, 1188–1206. https://doi.org/10.1016/j.apenergy.2017.07.035.
20. Gao, H.-C.; Choi, J.-H.; Yun, S.-Y.; Lee, H.-J.; Ahn, S.-J. Optimal scheduling and real-time control schemes of battery energy storage system for microgrids considering contract demand and forecast uncertainty. *Energies* **2018**, *11*, 1371. https://doi.org/10.3390/en11061371.
21. Shi, J.; Ye, Z.; Gao, H.O.; Yu, N. Lyapunov optimization in online battery energy storage system control for commercial buildings. *IEEE Trans. Smart Grid* **2023**, *14*, 328–340. https://doi.org/10.1109/TSG.2022.3197959.

22. Hasan, A.S.M.J.; Enriquez-Contreras, L.F.; Yusuf, J.; Ula, S. A universal optimization framework for commercial building loads using DERs from utility tariff perspective with tariff change impacts analysis. *Energy Rep.* **2023**, *9*, 6088–6101. https://doi.org/10.1016/j.egyr.2023.05.036.
23. Yu, M.G.; Huang, B.; Ma, X.; Devaprasad, K. Multi-objective sizing and dispatch for building thermal and battery storage towards economic and environmental synergy. *Appl. Energy* **2024**, *372*, 123819. https://doi.org/10.1016/j.apenergy.2024.123819.
24. Achour, Y.; Ouammi, A.; Zejli, D. Model predictive control based demand response scheme for peak demand reduction in a smart campus integrated microgrid. *IEEE Access* **2021**, *9*, 162765–162778. https://doi.org/10.1109/ACCESS.2021.3132895.
25. DTE Electric Company. *M.P.S.C. No. 1—Electric Rate Book, Rate Schedule D6.2: Primary Educational Institution Rate*; DTE Electric Company: Detroit, MI, USA, 2026.